\documentclass[prb,aps,twocolumn,superscriptaddress,showpacs]{revtex4}

\usepackage{amsmath}
\usepackage{amssymb}
\usepackage{graphicx}
\usepackage{bm}

\begin{document}

\bibliographystyle{apsrev}

\title{Quantum electrodynamics and photon-assisted tunnelling in long Josephson junctions}

\author{A.O. Sboychakov}
\affiliation{Frontier Research System, The Institute of Physical
and Chemical Research (RIKEN), Wako-shi, Saitama, 351-0198, Japan}
\affiliation{Institute for Theoretical and Applied Electrodynamics
Russian Academy of Sciences, Izhorskaya str. 13, Moscow 125412,
Russia}

\author{Sergey Savel'ev}
\affiliation{Frontier Research System, The Institute of Physical
and Chemical Research (RIKEN), Wako-shi, Saitama, 351-0198, Japan}
\affiliation{Department of Physics, Loughborough University,
Loughborough LE11 3TU, United Kingdom}

\author{Franco Nori}
\affiliation{Frontier Research System, The Institute of Physical
and Chemical Research (RIKEN), Wako-shi, Saitama, 351-0198, Japan}
\affiliation{MCTP, CSCS, Department of Physics, University of
Michigan, Ann Arbor, MI 48109, USA}

\begin{abstract}
We describe the interaction between an electromagnetic field and a
long Josephson junction (JJ) driven by a dc current. We calculate
the amplitudes of emission and absorption of light via the
creation and annihilation of quantized Josephson plasma waves
(JPWs). Both, the energies of JPW quanta and the amplitudes of
light absorption and emission, strongly depend on the junction's
length and can be tuned by an applied dc current. Moreover,
photon-assisted macroscopic quantum tunnelling in long Josephson
junctions show resonances when the frequency of the outside
radiation coincides with the current-driven eigenfrequencies of
the quantized JPWs.
\end{abstract}

\date{\today}

\maketitle

\section{Introduction}

The miniaturization of electronic devices allows the observation
of quantum effects which were impossible to measure in the past.
Systems of Josephson junctions, characterized by high frequency
(up to several THz), exhibit a crossover to the quantum regime at
relatively high temperatures. Indeed, quantum oscillations and
macroscopic quantum tunnelling~\cite{Rev1} (MQT) have been
observed in charge, flux, and phase qubits~\cite{pq}. Renewed
interest in MQT occurred after the recent discovery of MQT in high
temperature layered superconductors~\cite{JHTS1,JHTS2,JHTS3,Jin}.
The observed enhancement of MQT was attributed to the spatial
structure of the tunnelling fluxon~\cite{We,animation}. It is
important to develop a theory of quantum electrodynamics in long
(about $1$\,$\mu$m in stacks of JJs, and about tens of microns in
low-$T_c$ junctions) JJs where the spatial distribution of the
gauge-invariant phase difference is crucial. In this problem, the
standard quantum mechanical approach (where the phase difference
is associated with the coordinate of a quantum particle tunnelling
through an effective potential barrier) becomes invalid, and a
more advanced field-theoretical approach is
needed~\cite{weEPL,wePRB}.

Here we consider a Josephson junction (JJ) driven by a dc current
near its critical value and exposed to THz electromagnetic (EM)
waves. In this configuration, as known for point-like contacts,
the probability of MQT depends on the intensity and frequency
$\omega$ of the incident EM waves. In contrast to the
short-junction case, we predict several resonant enhancements of
the MQT escape rate, when the frequency $\omega$ matches the
eigenfrequencies of the JPWs. We also propose a full quantum
electrodynamical description of long JJs, to calculate the
probabilities of absorption and emission of light by JPW quanta.

In section II we derive the model and quantize the field of the
gauge-invariant phase difference $\varphi$. In section III we
consider the interaction of the quantized $\varphi$ with photons
and calculate the transition rates of absorbtion and emission of
light by JPW quanta. This allows us to find the mean values of
occupation numbers of JPW quanta and the mean energy of the
system, which is pumped by external THz radiation. In section IV
we calculate the probability of photon-assisted macroscopic
quantum tunnelling of the phase difference.

\section{Second quantization of the phase difference field}

\subsection{Lagrangian formulation}

\begin{figure}
\begin{center}
\includegraphics*[width=0.45\textwidth]{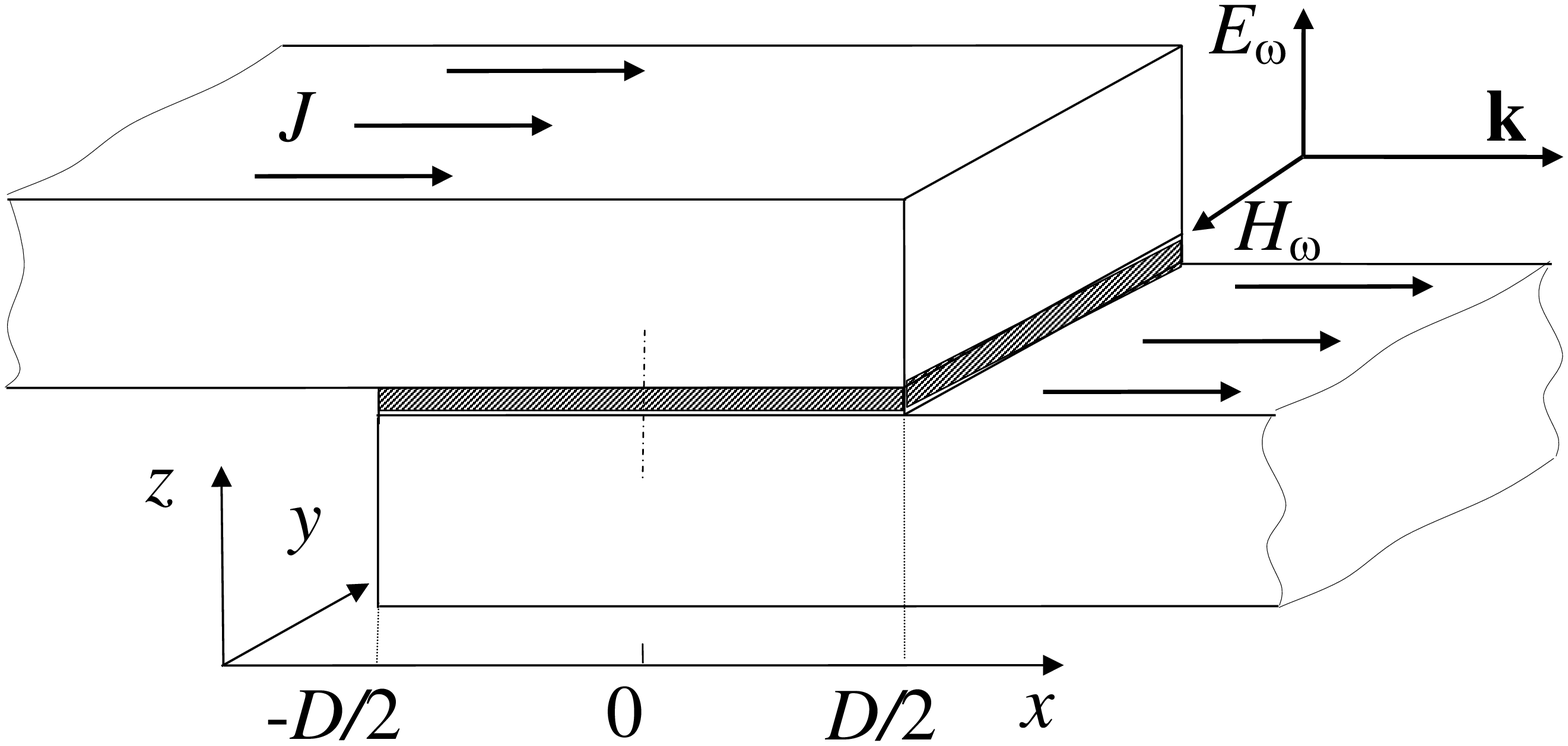}
\end{center}
\caption{\label{FigJJ} Schematic diagram of the Josephson
junction. The wave vector of the externally applied polarized THz
electromagnetic wave is directed along the $x$ axis, while its
electric (magnetic) field is directed along the $z$ ($y$) axis.}
\end{figure}

The geometry of the Josephson junction (JJ) under study is shown
in Fig.~\ref{FigJJ}. Two superconducting bars overlap a length $D$
in the $x$ direction. An insulating layer of thickness $s$, about
several nanometers, is placed between these two bars. A
supercurrent with density $i$ flows through the junction in the
$z$ direction. The width $L$ of the JJ in the $y$ direction is of
the order of, or less than, the Josephson penetration depth
$\lambda_J$, that is $l=L/\lambda_J\lesssim1$. The dynamics of the
gauge-invariant phase difference $\varphi(t,x,y)$ of such a
junction is described by the action
\begin{eqnarray}\label{S0} \
{\cal S}[\varphi]&=&\frac{1}{\omega_{p}}\int dt\Big({\cal L}[\varphi]+{\cal L}_{\Sigma}[\varphi]\Big),\nonumber\\
{\cal L}[\varphi]&=&\!\!\frac{\lambda_JE_J}{L}\int
dxdy\!\left[\frac12\left(\frac{\partial\varphi}{\partial
t}\right)^2-\frac12\left({\bm\nabla}\varphi\right)^2+\cos\varphi\right],\nonumber\\
{\cal L}_{\Sigma}[\varphi]&=&\frac{cE_J}{4\pi i_cL}\oint_{\Sigma}d\zeta\,\varphi\,\left[\mathbf{H}\times\mathbf{e}_{z}\right]_{n}\,.
\end{eqnarray}
In these equations, the $x$ and $y$ coordinates are normalized by
$\lambda_J$, the time $t$ is normalized by $1/\omega_{p}$, where
$\omega_{p}$ is the Josephson plasma frequency, $i_c$ is the
critical current density, and
$$
E_J=\hbar\omega_{p}{\Lambda}\,,\,\,\,\,{\Lambda}=\frac{i_c\lambda_JL}{2e\omega_{p}},
$$
where ${\Lambda}$ is considered to be much larger than unity,
${\Lambda}\gg1$. The integration in ${\cal L}_{\Sigma}[\varphi]$
is performed over a contour $\Sigma$ around the junction's area,
and the subscript $n$ refers to the component normal to the
contour $\Sigma$ in the $XY$ plane of the vector product of the
magnetic field $\mathbf{H}$ and unit vector $\mathbf{e}_z$.

The classical equation of motion for $\varphi$ with the
action~\eqref{S0} is the two dimensional Sine-Gordon equation
\begin{equation}\label{sinGE}
\frac{\partial^2\varphi}{\partial t^2}-\triangle\varphi+\sin\varphi=0\,.
\end{equation}
The surface term in the action~\eqref{S0} depicts the boundary
conditions to this equation,
\begin{equation}
\left.\frac{\partial\varphi}{\partial
n}\right|_{\mathbf{r}\in\Sigma}\!\!\!\!\!=[\mathbf{H}\times\mathbf{e}_{z}]_{n}.
\end{equation}
Representing the magnetic field in the form
$\mathbf{H}=\mathbf{H}_{J}+\mathbf{H}^{e}$, where $\mathbf{H}^{e}$
is the external ac magnetic field and $\mathbf{H}_{J}$ is the
field generated by the flowing current, we obtain
\begin{equation}\label{BoundConds}
\begin{array}{rcl}
\displaystyle\left.\frac{\partial\varphi}{\partial x}\right|_{x=\pm d/2}\!\!\!\!\!\!\!&=&\displaystyle\pm\frac{I}{2}+\frac{cH^{e}_{y}}{4\pi i_c\lambda_J}\,,\\
&&\\
\displaystyle\left.\frac{\partial\varphi}{\partial y}\right|_{y=\pm l/2}\!\!\!\!\!\!&=&\displaystyle-\frac{cH^{e}_{x}}{4\pi i_c\lambda_J}\,,
\end{array}
\end{equation}
where
$$
I=\frac{iD}{i_c\lambda_J},\,\,\,d=\frac{D}{\lambda_J},\,\,\,l=\frac{L}{\lambda_J}\,.
$$

When $\mathbf{H}^{e}=0$, the stationary solution to
Eq.~\eqref{sinGE}, corresponding to the lowest-energy minimum,
does not depend on the $y$-coordinate. Below we assume that the ac
magnetic field of the incident radiation is directed along the
$y$-axis (see Fig.~\ref{FigJJ}). In this case, only the plasma
waves in the $x$-direction are excited, and
$\varphi=\varphi(t,x)$. When $D\ll\lambda_J$, the field
$\varphi(t,x)$ only slightly depends on the $x$-coordinate, and
the action~\eqref{S0} describes the dynamics of the particle in
the washboard potential $V(\varphi)=-\cos\varphi-j\varphi$, where
$j=i/i_c$. When $j<1$, this potential has an infinite number of
minima, each one separated by a potential barrier of the order of
$\sqrt{1-j^2}$. The probability of quantum tunnelling from one
minimum to the nearest minimum can be easily calculated in the
semiclassical approximation~\cite{Caldeira}. When
$D\gtrsim\lambda_J$, the spatial dependence of the field
$\varphi(t,x)$ is essential, and the problem of quantum tunnelling
becomes more complicated. In the semiclassical approximation, the
probability of tunnelling can be written as~\cite{ColeI}
$$
\Gamma=\omega_0\sqrt{\frac{30B}{\pi}}\exp(-B)\,,
$$
where $B=2{\cal S}_{\text{E}}/\hbar$, and ${\cal S}_{\text{E}}$ is
the action, defined in Eq.~\eqref{S0}, in imaginary time $t=i\tau$
calculated along classical trajectories, and $\omega_0$ is the
oscillation frequency of the field $\varphi(t,x)$ near one of the
energy minima $\varphi_0(x)$. In one of our previous
papers~\cite{weEPL} we proposed an approach for calculating the
tunnelling exponent $B$ for a current $I=jd$ close to the critical
value $I_c(d)$ (which now nonlinearly depends on $d$). Here we
consider {\it the effect of external electromagnetic radiation} on
the probability of tunnelling.

\subsection{Quantum regime}

We consider the interaction of $\varphi$ with electromagnetic
waves by perturbations. First, we quantize the field
$\varphi(t,x)$ near the energy minimum $\varphi_0(x)$ at
$\mathbf{H}^{e}=0$, find the energy spectrum, and then calculate
the transition rates of the field $\varphi$, from the ground state
to its excited states and vice versa, due to the interaction with
the electromagnetic field. The knowledge of the transition rates
gives us the mean energy $\bar{E}(\omega,P)$ of the field
$\varphi$ in the presence of an external radiation as a function
of its power $P$ and frequency $\omega$. Since the effective
potential barrier decreases with the growth of $\bar{E}$, the
external radiation enhances the tunnelling. It is clear that a
strong enhancement of the escape rate $\Gamma$ should be at
frequencies $\omega$ close to the eigenfrequencies $\omega_n$ of
the $\varphi$ field. The tunnelling exponent $B$, as function of
$\bar{E}$, is found here using the approach described in
Ref.~\onlinecite{weEPL}.

The static solution corresponding to an energy minimum satisfies
the static Sine-Gordon equation
\begin{equation}
\frac{d^2\varphi_0}{dx^2}=\sin\varphi_0
\end{equation}
with the boundary conditions
\begin{equation}
\left.\frac{d\varphi_0}{dx}\right|_{x=\pm
d/2}\!\!\!=\,\pm\,\frac{I}{2}\,.
\end{equation}
The solution to this equation exists for currents $I$ less than
the critical value $I_c(d)$. If $d\lesssim 4\lambda_J$, the
current density in the JJ is approximately constant and the
function $I_c(d)$ increases linearly with $d$; if $d\gg
\lambda_J$, the current flows near the junction edges and $I_c(d)$
reaches the saturation value $I_c^{\text{max}}=4$. In order to
quantize $\varphi$ we represent it in the form
$$
\hat{\varphi}(t,x)=\varphi_0(x)+\hat{\psi}(t,x)\,,
$$
where the operator $\hat{\psi}$ satisfies the boundary conditions
\begin{equation}
\left.\frac{d\hat{\psi}}{dx}\right|_{x=\pm d/2}\!\!\!\!=0\,,
\end{equation}
and expand the Lagrangian ${\cal L}$ in powers of $\hat{\psi}$. We
introduce the momentum
\begin{equation}\label{Pi}
\hat{\pi}(t,x)=\frac{\delta{\cal
L}}{\delta\left(\frac{\partial\hat{\psi}(t,x)}{\partial
t}\right)}=\hbar\omega_{p}\,\Lambda\,\frac{\partial\hat{\psi}(t,x)}{\partial
t},
\end{equation}
and require the standard simultaneous commutation relation
\begin{equation}\label{ComRel}
\left[\hat{\psi}(t,x),\hat{\pi}(t,x')\right]_{-}=i\hbar\delta(x-x')\,.
\end{equation}
The Hamiltonian of the system, $\hat{\cal H}$, has a form
\begin{equation}
\hat{{\cal H}}=\hat{\pi}\frac{\partial\hat{\psi}}{\partial
t}-\hat{{\cal L}}=\hat{{\cal H}}_0+\hat{{\cal H}}'\,,
\end{equation}
where
\begin{eqnarray}\label{H0int}
\hat{{\cal
H}}_0&=&E_J:\!\!\int_{-d/2}^{d/2}\!\!\!\!dx\left[\frac{1}{2E_J^2}\hat{\pi}^2+\frac12\hat{\psi}\hat{\cal D}\hat{\psi}\right]:\,,\\
\hat{{\cal
H}}'&=&E_J:\!\!\int_{-d/2}^{d/2}\!\!\!\!dx\left[\frac16\sin\varphi_0\,\hat{\psi}^3+\frac{1}{24}\cos\varphi_0\,\hat{\psi}^4+\ldots\right]:\,.\nonumber
\end{eqnarray}
In these equations, the colons ``\,:\,'' mean ``normal ordering''
and $\hat{\cal D}$ is a differential operator of the form
\begin{equation}
\hat{\cal D}=-\frac{\partial^2}{\partial x^2}+\cos(\varphi_0(x))\,.
\end{equation}
In the interaction representation, the operators $\hat{\psi}$ and
$\hat{\pi}$ can be written as
\begin{eqnarray}\label{PsiPiExp}
\hat{\psi}&=&\sqrt{\frac{1}{2{\Lambda}}}\sum_n\frac{\psi_n(x)}{\mu_n^{1/4}}\left(\text{e}^{i\sqrt{\mu_n}\,t}\,\hat{b}_n^{\dag}+%
\text{e}^{-i\sqrt{\mu_n}\,t}\,\hat{b}_n\right)\!,\\
\hat{\pi}&=&i\hbar\omega_{p}\sqrt{\frac{{\Lambda}}{2}}\sum_n\frac{\psi_n(x)}{\mu_n^{1/4}}\left(\text{e}^{i\sqrt{\mu_n}\,t}\,\hat{b}_n^{\dag}-%
\text{e}^{-i\sqrt{\mu_n}\,t}\,\hat{b}_n\right)\!,\nonumber
\end{eqnarray}
where $\mu_n$, $\psi_n$ are, respectively, the eigenvalues and
orthogonal eigenfunctions of the operator $\hat{\cal D}$, that is,
\begin{equation}\label{eigenf}
\hat{\cal
D}\psi_n=\mu_n\psi_n\,,\,\,\,\,\int_{-d/2}^{d/2}\!\!\!\!dx\,\psi_n(x)\psi_m(x)=\delta_{nm}\,,
\end{equation}
In Eq.~\eqref{PsiPiExp}, $\hat{b}_n^{\dag}$ and $\hat{b}_n$ are
the creation and annihilation operators of JPW quanta in the state
$n$. Note that all $\mu_n$'s are positive when $I<I_c(d)$ because
the $\varphi_0(x)$ corresponds to an energy minimum. In terms of
the operators $\hat{b}_n^{\dag}$ and $\hat{b}_n$, the Hamiltonian
$\hat{{\cal H}}_0$ takes the form
\begin{equation}
\hat{{\cal
H}}_0=\hbar\omega_{p}\sum_n\sqrt{\mu_n}\;\hat{b}_n^{\dag}\hat{b}_n\,.
\end{equation}
The Hamiltonian $\hat{{\cal H}}'$ describes the self-interaction
of the field $\varphi$. Since ${\Lambda}\gg1$, $\hat{{\cal H}}'$
can be considered as a perturbation if the energy of the system
(counting from the `vacuum' state corresponding to $\varphi_0(x)$)
$\bar{E}\ll\hbar\omega_{p}{\Lambda}$. In zeroth order, this energy
is determined by the occupation numbers $N_n$, and reads
$$
\bar{E}=\hbar\omega_{p}\sum_n\sqrt{\mu_n}\,N_n\,.
$$
The correction to this result due to self interactions can be
found via perturbation theory.

\section{Interaction with an electromagnetic field: Absorption and emission transition rates}

Now we consider the interaction of the field $\varphi$ with
electromagnetic waves, described by the vector potential
$\mathbf{A}$ (we choose the gauge $A_0=0,\,{\rm
div}\mathbf{A}=0$). Substituting $\varphi=\varphi_0+\psi$ into the
action~\eqref{S0} and expanding it in a power series of $\psi$, we
derive the operator $\hat{V}$ describing the interaction of $\psi$
with the electromagnetic field~\footnote{The action~\eqref{S0} has
also a surface term, proportional to
$\oint_{\Sigma}d\zeta\,\varphi_0\left[{\rm
rot}\mathbf{A}\times\mathbf{e}_{z}\right]_{n}$. Since we consider
an ac magnetic field, the time-average of this term is zero, and
we neglect it. }:
\begin{equation}\label{V0}
\hat{V}=-\frac{cE_J}{4\pi i_c\lambda_JL}\oint_{\Sigma}d\zeta\,\hat{\psi}\,\left[{\rm rot}\mathbf{A}\times\mathbf{e}_{z}\right]_{n}\,.
\end{equation}
Here we use the relation $\mathbf{H}^{e}={\rm
rot}\mathbf{A}/\lambda_J$ because we measure distances in units of
$\lambda_J$. The vector potential $\mathbf{A}$ in Eq.~\eqref{V0}
consists of two parts, describing both the incoming and outgoing
radiation re-emitted by the JJ.  We assume that the incident
electromagnetic radiation is fully polarized and propagates along
the $x$ axis, as shown in Fig.~\ref{FigJJ}. Below we measure the
frequency $\omega$ in units of $\omega_{p}$ and the wave length in
units of $\lambda_J$. In this case, $\mathbf{A}$ can be written as
\begin{eqnarray}\label{A}
\hat{\mathbf{A}}(t,\mathbf{r})&=&-i\mathbf{e}_z\frac{c}{\omega_p}\int\frac{d\omega}{2\pi}\frac{E_{\omega}}{\omega}\text{e}^{-i\omega(t-vx)}+\nonumber\\
&&\sqrt{\frac{4\pi c^2}{V\omega_{p}}}\sum_{\mathbf{k},\lambda}%
\left(\sqrt{\frac{\hbar}{2\omega_{\mathbf{k}}}}\mathbf{e}^{\lambda}(\mathbf{k})\text{e}^{-i\omega_{\mathbf{k}}t+i\mathbf{kr}}\hat{a}_{\mathbf{k}\lambda}+\right.\nonumber\\
&&\left.\sqrt{\frac{\hbar}{2\omega_{\mathbf{k}}}}\mathbf{e}^{\lambda}(\mathbf{k})\text{e}^{i\omega_{\mathbf{k}}t-i\mathbf{kr}}\hat{a}_{\mathbf{k}\lambda}^{\dag}\right),
\end{eqnarray}
where $\hat{a}_{\mathbf{k}\lambda}^{\dag}$ and
$\hat{a}_{\mathbf{k}\lambda}$ are the creation and annihilation
operators of a photon with wave vector $\mathbf{k}$ and
polarization $\lambda$,
$$
v=\frac{\omega_{p}\lambda_J}{c}
$$
is the ratio of the Swihart velocity $\omega_p\lambda_J$ to the
speed of light $c$, $\omega_{\mathbf{k}}=|\mathbf{k}|/v$, and $V$
is the volume of space (dimensional) where the electromagnetic
field exists. The first term in Eq.~\eqref{A}  corresponds to
incoming radiation (which is here considered as classical), where
$E_{\omega}$ is the electric field at frequency $\omega$. The
second term describes the photons appearing due to the interaction
of the incoming electromagnetic waves with the JJ. In this term,
$\mathbf{e}^{\lambda}$ is the vector of polarization, which
satisfies the equality
\begin{equation}\label{pol}
\mathbf{k}\cdot\mathbf{e}^{\lambda}(\mathbf{k})=0\,.
\end{equation}
Substituting Eq.~\eqref{A} and the expansion~\eqref{PsiPiExp} for
$\hat{\psi}$ into Eq.~\eqref{V0},  and performing the surface
integration, we derive:
$$
\hat{V}=\hat{V}_{\text{ext}}+\hat{V}_{\text{q}}\,,
$$
\begin{eqnarray}
\hat{V}_{\text{ext}}&=&i\sqrt{\frac{\hbar c\lambda_LLv}{4\pi}}\sum_n\int\frac{d\omega}{2\pi}E_{\omega}%
\frac{\varkappa_n(\omega v)}{\sqrt{\omega_n}}\text{e}^{-i\omega
t}\times\nonumber\\
&&\left(\text{e}^{i\omega_nt}\hat{b}_n^{\dag}+\text{e}^{-i\omega_nt}\hat{b}_n\right)\label{VExt}\\
\hat{V}_{\text{q}}&=&\hbar c\sqrt{\frac{\lambda_LL}{2V\lambda_J}}\sum_n\sum_{\mathbf{k}\lambda}%
\frac{S(k_y)}{\sqrt{\omega_{\mathbf{k}}\omega_n}}\times\nonumber\\
&&\Big(\!\!\left[k_yh^{\lambda}_x(\mathbf{k})\chi_n(k_x)+h^{\lambda}_y(\mathbf{k})\varkappa_n(k_x)\right]%
\text{e}^{-i\omega_{\mathbf{k}}t}\hat{a}_{\mathbf{k}\lambda}+\nonumber\\
&&\left[k_yh^{\lambda}_x(\mathbf{k})\chi_n^{*}(k_x)+h^{\lambda}_y(\mathbf{k})\varkappa_n^{*}(k_x)\right]%
\text{e}^{i\omega_{\mathbf{k}}t}\hat{a}_{\mathbf{k}\lambda}^{\dag}\Big)\times\nonumber\\
&&\left(\text{e}^{i\omega_nt}\hat{b}_n^{\dag}+\text{e}^{-i\omega_nt}\hat{b}_n\right)\,,\label{VQ}
\end{eqnarray}
where $\lambda_L$ is the London penetration depth, which is linked
to $i_c$ and $\lambda_J$ by the relation (it is supposed here that
$s\ll\lambda_L$)
$$
\lambda_J^2=\frac{\hbar c^2}{8\pi e(2\lambda_L+s)i_c}\cong\frac{\hbar c^2}{16\pi e\lambda_Li_c}.
$$
In formulas~\eqref{VExt} and~\eqref{VQ}
\begin{equation}
\omega_n=\sqrt{\mu_n}\,,
\end{equation}
\begin{equation}
\mathbf{h}^{\lambda}(\mathbf{k})=\mathbf{k}\times\mathbf{e}^{\lambda}(\mathbf{k})\,,
\end{equation}
and functions $\varkappa_n(k)$, $\chi_n(k)$, and $S(k)$ are the following:
\begin{equation}\label{kappa}
\varkappa_n(k)=-i\left[\psi_n\big(d/2\big)\text{e}^{ikd/2}-\psi_n\big(-d/2\big)\text{e}^{-ikd/2}\right],
\end{equation}
\begin{equation}\label{chi}
\chi_n(k)=\int_{-d/2}^{d/2}\!\!\!\!dx\,\psi_n(x)\text{e}^{ikx},
\end{equation}
\begin{equation}\label{Sy}
S(k)=\frac{2\sin(kl/2)}{kl}\,.
\end{equation}

\subsection{Spontaneous photon emission}

In first order of perturbation theory, there are only three
possible processes: (i) spontaneous emission of a photon by the
field $\varphi$, (ii) induced photon absorption, and (iii) induced
emission~\footnote{We do not consider the process of spontaneous
absorption since the intensity of re-emitted photons by the
junction is assumed to be negligible in comparison to the
intensity of incoming radiation.}. Let us first consider the {\it
spontaneous} emission, which is described by the operator
$\hat{V}_q$. In the initial state, we have the set of occupation
numbers of JPW quanta, $\{N_m\}$, and zero photons, and, in the
final state, one of these numbers, say $N_n$, decreases by $1$ and
one photon appears in the system. We neglect the effect of thermal
radiation, proceeding to the limit $T\to0$. The probability per
unit time of such a process, $w^{(-)}$, is proportional to $N_n$.

Following the standard approach~\cite{LL4}, we derive for the
probability of emission of a photon with wave vector $\mathbf{k}$
and polarization $\lambda$:
\begin{equation}\label{dwm}
dw^{(-)}_{n\lambda}(\mathbf{k})\!=N_n\frac{\pi c\lambda_LL}{v\lambda_J^3}%
\frac{\left|F^{\lambda}_n(\mathbf{k})\right|^2}{\omega_{\mathbf{k}}\,\omega_n}\,\delta(\omega_{\mathbf{k}}-\omega_n)%
\frac{d^3\mathbf{k}}{(2\pi)^3},
\end{equation}
where
\begin{equation}
F^{\lambda}_n(\mathbf{k})=S(k_y)\left[k_yh^{\lambda}_x(\mathbf{k})\chi_n(k_x)%
+h^{\lambda}_y(\mathbf{k})\varkappa_n(k_x)\right].
\end{equation}
We introduce spherical coordinates in momentum space. Performing
the integration over $\mathbf{k}$ and  the summation over
$\lambda$ taking into account the relation~\eqref{pol}, finally,
we derive
\begin{equation}\label{wm}
w^{(-)}_n=N_n\omega_{p}\gamma_n,\,\,\,\,\gamma_n=\frac{\lambda_LLv}{2\pi\lambda_J^2}\,\nu_n\,,
\end{equation}
\begin{equation}\label{nu}
\nu_n=\frac{1}{4\pi}\sum_{\lambda}\int\!\!d\mathbf{m}\left|F^{\lambda}_n(\omega_{n}v\mathbf{m})\right|^2\,,
\end{equation}
where $\mathbf{m}$ is a unit vector in momentum space.

For not too long junctions, it is possible to obtain an analytical
expression for $\nu_n$. The wave length $\lambda$ of the
electromagnetic radiation under consideration is about
$$
\lambda\sim c/\omega_p=\lambda_J/v\gg\lambda_J\,,
$$
since the typical value of $v\sim3\cdot10^{-2}\ll1$. Therefore,
for $d\ll\lambda$, one can expand
$F^{\lambda}_n(\omega_{n}v\mathbf{m})$ in Eq.~\eqref{nu} in powers
of $v$. Doing so, we derive in the lowest order
\begin{eqnarray}\label{nuappr}
\nu_{2m+1}&\cong&\!\!\frac{8}{3}\omega_{2m+1}^2v^2\,\psi_{2m+1}^2\Big(\frac{d}{2}\Big)\,,\\
\nu_{2m}&\cong&\!\!\frac{4d^{2}}{15}\omega_{2m}^4v^4\!\!%
\left[\psi_{2m}^2\Big(\frac{d}{2}\Big)-\frac12\bar{\psi}_{2m}\psi_{2m}\Big(\frac{d}{2}\Big)+\bar{\psi}_{2m}^2\right]\!,\nonumber
\end{eqnarray}
where
\begin{equation}
\bar{\psi}_{n}=\frac{1}{d}\int_{-d/2}^{d/2}\!\!\!\!dx\,\psi_n(x)
\end{equation}
and $\nu_{2m}\ll\nu_{2m+1}$. The difference between $\nu_n$ with
odd and even $n$ comes from the symmetry properties of the JPW
wave functions: $\psi_n(-x)=(-1)^n\psi_n(x)$.

The value of $\gamma_n$ gives us the radiation width of the $n$th
level in units of $\omega_{p}$. When $d\sim\lambda_J$, we have
from~\eqref{nuappr} $\nu_{2m+1}\sim v^2$ and $\nu_{2m}\sim v^4$.
Considering $L,\lambda_J\sim10^{-3}\,$cm,
$\lambda_L\sim10^{-5}\,$cm, and $v\sim3\cdot10^{-2}$ we obtain
$\gamma_{2m+1}\sim10^{-8}$--$10^{-7}$ and
$\gamma_{2m}\sim10^{-12}$--$10^{-10}$. Note that we do not
consider here another possible mechanisms of dissipation, which
can substantially increase the width of the JPW quanta energy
levels.

\subsection{Induced photon absorbtion and emission}

Let us now consider processes of {\it induced} photon absorption
and emission. These two processes are determined by the operator
$\hat{V}_{\text{ext}}$. We denote by $^{\text{in}}w^{(+)}_n$
($^{\text{in}}w^{(-)}_n$) the probability per unit time of
creation (annihilation) of a quantum of the $\varphi$ field in the
$n$th state due to induced photon absorption (emission). These two
probabilities satisfy the following equality
$$
\frac{^{\text{in}}w^{(+)}_n}{^{\text{in}}w^{(-)}_n}=\frac{N_n+1}{N_n}\,.
$$
Thus, to first order in perturbation theory, the probability per
unit time of induced photon absorption and also accounting for
induced emission,
$w^{(+)}_n=^{\text{in}}w^{(+)}_n-^{\text{in}}w^{(-)}_n$, does not
depend on $N_n$, and is only determined by the power and frequency
of the external radiation. Making a similar calculation as for
$w^{(-)}_n$, we derive
\begin{equation}\label{wp0}
w^{(+)}_n=\frac{c\lambda_LLv}{2\hbar\omega_{p}}\int%
d\omega\,\frac{\left|E_{\omega}\right|^2}{2\pi}\frac{\left|\varkappa_n(\omega v)\right|^2}{\omega_n}\,\delta(\omega-\omega_n)\,.
\end{equation}
We assume that the incident radiation has a Gaussian distribution
with central frequency $\bar{\omega}$ and width $\bar{\gamma}$,
that is
\begin{equation}
\frac{\left|E_{\omega}\right|^2}{2\pi}=\frac{4\pi
P}{c}\Gamma(\omega-\bar{\omega}),
\end{equation}
where $P$ is the radiation power per unit area and
\begin{equation}\label{Gamma}
\Gamma(\omega)=\frac{1}{\bar{\gamma}\sqrt{\pi}}\,\exp\left\{-\omega^2/\bar{\gamma}^2\right\}.
\end{equation}
The probability $w^{(+)}_n$ then becomes
\begin{equation}\label{wp}
w^{(+)}_n=\frac{2\pi P\lambda_LLv}{\hbar\omega_{p}}\frac{f_n(\bar{\omega})}{\omega_n}\,,
\end{equation}
where
\begin{equation}\label{f}
f_n(\bar{\omega})=\int d\omega\,\Gamma(\omega-\bar{\omega})\left|\varkappa_n(\omega v)\right|^2\delta(\omega-\omega_n).
\end{equation}

In equilibrium, the probabilities $w^{(+)}_n$ and $w^{(-)}_n$
coincide. This gives rise to a relation for the mean values of the
occupation numbers $\bar{N}_n$:
\begin{equation}\label{meanN}
\bar{N}_n=\frac{4\pi^2P\lambda_J^2}{\hbar\omega_{p}^2}\frac{f_n(\bar{\omega})}{\omega_n\nu_n}.
\end{equation}
The mean value $\bar{E}$ of the system energy (to zeroth order in
$\hat{\cal H}'$) then reads
\begin{equation}\label{E}
\bar{E}=\frac{4\pi^2P\lambda_J^2}{\omega_{p}}\sum_n\frac{f_n(\bar{\omega})}{\nu_n}.
\end{equation}

If the frequency band of the radiation source is large enough,
that is, $\bar{\gamma}\gg\gamma_n$, we can easily perform an
integration in Eq.~\eqref{f}. As a result, the mean energy becomes
\begin{equation}\label{EGauss}
\bar{E}=\frac{4\pi^2P\lambda_J^2}{\omega_{p}}\sum_n\frac{\left|\varkappa_n(\omega_nv)\right|^2}{\nu_n}\Gamma(\bar{\omega}-\omega_n).
\end{equation}
In the opposite case of near-monochromatic radiation,
$\bar{\gamma}\ll\gamma_n$, we should take into account that the
energy levels of JPW quanta have finite width $\gamma_n$ (in units
of $\omega_{p}$). Replacing the delta function in Eq.~\eqref{f} by
$$
\delta(\omega-\omega_n)\rightarrow\frac{1}{\pi}\frac{\gamma_n}{(\omega-\omega_n)^2+\gamma_n^2}
$$
and using
$\Gamma(\omega-\bar{\omega})=\delta(\omega-\bar{\omega})$, we
obtain
\begin{equation}\label{ELorenz}
\bar{E}=\frac{2P\lambda_LLv}{\omega_{p}}\sum_n\frac{\left|\varkappa_n(\omega_nv)\right|^2}{(\bar{\omega}-\omega_n)^2+\gamma_n^2},
\end{equation}
where we take into account that $\gamma_n\ll1$. Let us notice that
formulas~\eqref{E}, \eqref{EGauss}, and~\eqref{ELorenz} are valid
only at not too high radiation power when
$\bar{E}\ll\hbar\omega_{p}{\Lambda}$.

\subsection{Response of the junction to a wave packet}

\begin{figure}
\begin{center}
\includegraphics*[width=0.45\textwidth]{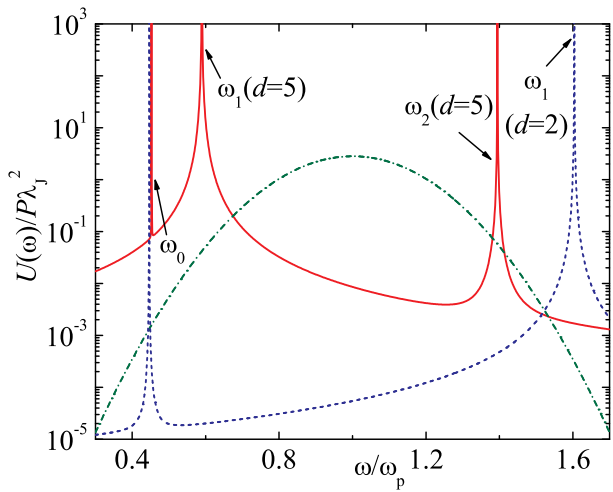}
\end{center}
\caption{\label{FigU} (Color online) The frequency dependence of
the intensity of re-emission, $U(\omega)$, calculated for $d=5$
(red solid curve) and $d=2$ (blue dashed curve). Other parameters
are: $I/I_c(d)=0.98$, $\lambda_L/\lambda_J=4\cdot10^{-3}$, and
$L/\lambda_J=2$. The green dot-dashed curve corresponds to
Gaussian distribution of the intensity of incoming radiation with
central frequency $\bar{\omega}=1$ and the width
$\bar{\gamma}=0.2$ (in units of $\omega_p$).}
\end{figure}

Consider now the response of a JJ to a wide-band THz wave packet.
We now assume that the central frequency of the incoming radiation
$\bar{\omega}$ is about $\omega_p$ and that the width
$\bar{\gamma}$ of the wave packet is large enough. In this case,
the first several energy levels of the system will be excited. The
intensity $U(\omega)$ of light re-emission at frequency $\omega$
is given by the sum
$\hbar\omega_p\sum_{n\lambda}\omega_ndw^{(-)}_{n\lambda}(\mathbf{k})$,
with $dw^{(-)}_{n\lambda}(\mathbf{k})$ from Eq.~\eqref{dwm},
integrated over all directions of $\mathbf{k}$. Taking into
account the relation~\eqref{meanN} for the mean values of the
occupation numbers $\bar{N}_n$, and replacing again the delta
function in Eq.~\eqref{dwm} by a Lorentzian curve, we obtain
\begin{equation}
U(\omega)=2P\lambda_LL\sum_n\frac{\left|\varkappa(\omega_nv)\right|^2\Gamma(\bar{\omega}-\omega_n)}%
{(\omega-\omega_n)^2+\gamma_n^2}\,.
\end{equation}
The function $U(\omega)$, for relatively short ($d=2$) and long
JJs ($d=5$), is shown in Fig.~\ref{FigU}. The wave packet of the
incident radiation has a central frequency $\bar{\omega}=1$ and
width $\bar{\gamma}=0.2$ (in units of $\omega_p$). In this case,
the first two ($d=2$) or three ($d=5$) energy levels are excited.
For short junctions, $d\lesssim1$, the eigenfrequencies are,
approximately
\begin{equation}
\omega_0\approx(1-j^2)^{1/4}\,,\,\,\,\,\omega_n\approx\frac{\pi n
}{d}\gg\omega_0\,,\,\,\,n>0\,.
\end{equation}
For increasing values of $d$, $\omega_1$ tends to $\omega_0$, and
when $d\gtrsim4$, we have
\begin{equation}
\omega_0\approx\omega_1\ll\omega_n\,,\,\,\,n>1\,.
\end{equation}
The relation $\omega_1\approx\omega_0$ is essential for the
properties of macroscopic quantum tunnelling in JJs. Namely, in
this case we have two channels of tunnelling, corresponding to
fluxons arising near the junction's edges. This situation is
considered in the next section.

\section{Probability of photon-assisted tunnelling}

Now we calculate the probability per unit time of quantum
tunnelling to another vacuum state, stimulated by external
electromagnetic radiation, using the approach~\cite{weEPL}
proposed in one of our previous papers. In the semiclassical
approximation, we can consider the quantum field $\hat{\psi}$ as a
classical field $\psi(\tau,x)$ in imaginary time $t=i\tau$. The
probability, $\Gamma$, then reads~\cite{ColeI}
\begin{equation}\label{GammaE}
\Gamma(\bar{E})=\omega_{p}\sqrt{\frac{30B(\bar{E})\mu_0}{\pi}}\exp[-B(\bar{E})],\,\,\,\,B(\bar{E})=\frac{2S_E}{\hbar}\,,
\end{equation}
where ${\cal S}_E$ is the action~\eqref{S0} in imaginary time.
Substituting
$$
\varphi(\tau,x)=\varphi_0(x)+\psi(\tau,x)
$$
into Eq.~\eqref{S0} and expanding the action in powers of $\psi$,
we obtain
\begin{eqnarray}\label{BE}
\!\!\!\!\!\!\!\!\!B(\bar{E})&=&%
\!\!2{\Lambda}\int_{0}^{\tau_0}\!\!\!\!d\tau\!\!\left\{\int_{-d/2}^{d/2}\!\!\!\!\!\!\!\!dx\left[\frac12\psi\left(\hat{\cal D}-\frac{\partial^2}{\partial\tau^2}\right)\right.\right.\psi-\nonumber\\%
&&\!\!\!\!\left.\left.\frac16\sin\varphi_0\,\psi^3-\frac{1}{24}\cos\varphi_0\,\psi^4-\dots\right]-\frac{\bar{E}}{E_J}\right\}.
\end{eqnarray}
The last term in Eq.~\eqref{BE} originates from the matching
condition for the wave function $\Phi$ of the quantum field
$\hat{\psi}$ inside ($\Phi_{\text{in}}\propto\exp(-{\cal
S}_E/\hbar)$) and outside
($\Phi_{\text{out}}\propto\exp(-i\bar{E}t/\hbar)$) the barrier.
The field $\psi$ in Eq.~\eqref{BE} satisfies the equation $\delta
B(\bar{E})/\delta\psi=0$, that is
\begin{equation}\label{EqPsi}
\frac{\partial^2\psi}{\partial\tau^2}-\hat{\cal D}\psi=%
-\frac12\sin\varphi_0\,\psi^2-\frac{1}{6}\cos\varphi_0\,\psi^3-\dots\,,
\end{equation}
with the following initial and boundary conditions
$$
\left.\frac{\partial\psi}{\partial\tau}\right|_{\tau=0,\tau_0}\!\!\!\!\!\!=0,\,\,\,\,%
\left.\frac{\partial\psi}{\partial x}\right|_{x=\pm
d/2}\!\!\!\!\!\!=0\,.
$$

We seek a solution of the equation~\eqref{EqPsi} in the form
\begin{equation}
\psi(\tau,x)=\sum_nc_n(\tau)\,\psi_n(x)\,.
\end{equation}
Multiplying Eq.~\eqref{EqPsi} by $\psi_n$ and performing space
integration and using Eq.~\eqref{eigenf}, we obtain the system of
equations for $c_n(\tau)$
\begin{equation}\label{eqC}
\ddot{c}_n-\mu_nc_n=-\frac12\sum_{mk}U^{(3)}_{nmk}\,c_mc_k
-\frac16\sum_{mkl}U^{(4)}_{nmkl}\,c_mc_kc_l-\dots
\end{equation}
with initial conditions
\begin{equation}\label{init}
\dot{c}_n(0)=\dot{c}_n(\tau_0)=0\,.
\end{equation}
Here, the dot means ``imaginary time derivative'', and
\begin{equation}
U^{(i)}_{n\;...\;
k}=-\int_{-d/2}^{d/2}\!\!\!dx\,\frac{\partial^i(\cos\varphi_0)}{\partial\varphi_0^i}\,\psi_n\,...\,\psi_k.
\end{equation}
The tunnelling exponent $B(\bar{E})$, Eq.~\eqref{BE}, can be
expressed as
\begin{eqnarray}\label{BEcn}
\!B(\bar{E})&=&{\Lambda}\!\int_{0}^{\tau_0}\!\!\!d\tau\!%
\left[\,\frac16\sum_{nmk}U^{(3)}_{nmk}\,c_nc_mc_k-\frac{2\bar{E}}{E_J}\,+\right.\nonumber\\
&&\left.\frac{1}{12}\sum_{nmkl}U^{(4)}_{nmkl}\,c_nc_mc_kc_l+\dots\right].
\end{eqnarray}

When the current $I$ is close to the critical value $I_c(d)$, we
have $\mu_0\ll1$ and $c_n\ll1$. So, we can neglect all terms in
the right-hand-side of Eq.~\eqref{eqC}, except the first one. Our
analysis shows that when $d\gtrsim4$, $\mu_1\approx\mu_0$, and we
have the following relation for the eigenvalues of the operator
$\hat{\cal D}$
$$
\mu_0\approx\mu_1\ll\mu_n\,,\,\,\,n>1,
$$
In this case, $c_0,c_1\ll c_n$ ($n>1$), and we can consider only
the first two equations of the system~\eqref{eqC}, taking $c_n=0$
for all $n>1$ (for details, see Ref.~\onlinecite{weEPL}).

\begin{figure}
\begin{center}
\includegraphics*[width=0.45\textwidth]{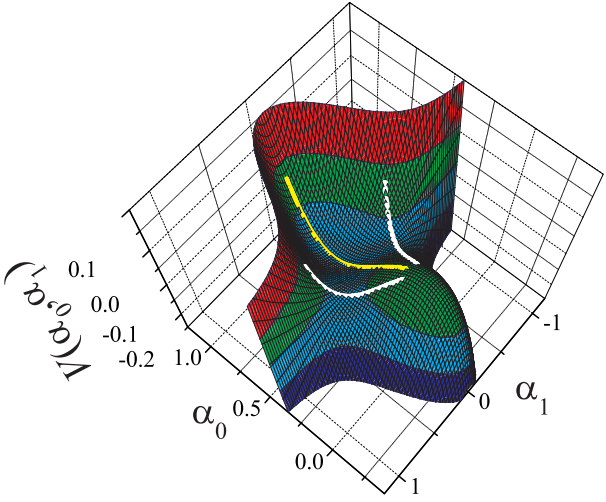}
\end{center}
\caption{\label{FigV} (Color online) The potential
$V(\alpha_0,\alpha_1)$ calculated when $d=6$, $I/I_c(d)=0.98$
($\lambda/u_{01}=1.18$). A particle tunnels from its initial
position near $\alpha_i=0$. The $\alpha_i$ are collective
coordinates for the tunnelling fluxon. The curves correspond to
three possible imaginary-time trajectories of the particle:
$\alpha_i^{(0)}(\eta)$ and $\alpha_i^{(\pm)}(\eta)$ (see text
below). Note, that the real-time potential equals to
$-V(\alpha_0,\alpha_1)$.}
\end{figure}

We now introduce new variables
\begin{equation}\label{alpha}
\alpha_i(\eta)=\frac{\sqrt{u_0u_i}\,c_i(\tau)}{3\mu_0},\,\,\,i=0,1\,,
\end{equation}
where
\begin{equation}
\eta=\sqrt{\mu_0}\tau\,,\,\,\,u_i=U^{(3)}_{0ii}.
\end{equation}
The system of equations~\eqref{eqC} takes the form
\begin{equation}\label{eqAlpha}
\left\{\begin{array}{rcl}
\displaystyle\frac{d^2\alpha_0}{d\eta^2}-\alpha_0&=&-\displaystyle\frac32\left(\alpha_0^2+\alpha_1^2\right)\,,\\
\\\displaystyle\frac{d^2\alpha_1}{d\eta^2}-\lambda\,\alpha_1&=&-3u_{01}\,\alpha_0\alpha_1\,,
\end{array}\right.
\end{equation}
where
\begin{equation}
\lambda=\frac{\mu_1}{\mu_0}\,,\,\,\,u_{01}=\frac{u_1}{u_0}.
\end{equation}
The system~\eqref{eqAlpha} has the first integral
\begin{equation}\label{FirstInt}
\frac{9\mu_0^3}{u_0^2}\left[\left(\frac{d\alpha_0}{d\eta}\right)^2+\frac{1}{u_{01}}\left(\frac{d\alpha_1}{d\eta}\right)^2+V(\alpha_0,\alpha_1)\right]=-\frac{\bar{E}}{E_J},
\end{equation}
where we introduce a potential
\begin{equation}\label{Potential}
V(\alpha_0,\alpha_1)=\alpha_0^3+3\alpha_0\alpha_1^2-\alpha_0^2-\frac{\lambda}{u_{01}}\,\alpha_1^2\,.
\end{equation}
Taking into account the initial conditions~\eqref{init}, we have,
at the turning points:
\begin{equation}\label{tauE}
\left.V\left(\alpha_0(\sqrt{\mu_0}\,\tau),\,\alpha_1(\sqrt{\mu_0}\,\tau)\right)\right|_{\tau=0,\tau_0}\!\!=-\bar{\varepsilon}\,,
\end{equation}
where
\begin{equation}\label{epsilon}
\bar{\varepsilon}=\frac{u_0^2\bar{E}}{9\mu_0^3E_J}\,,\,\,\,\,\,\,\,\,0<\bar{\varepsilon}<\varepsilon_0=\frac{4}{27}\,.
\end{equation}
The equation~\eqref{tauE} defines the value of $\tau_0$ as a
function of system energy $\bar{E}$.

\begin{figure}
\begin{center}
\includegraphics*[width=0.45\textwidth]{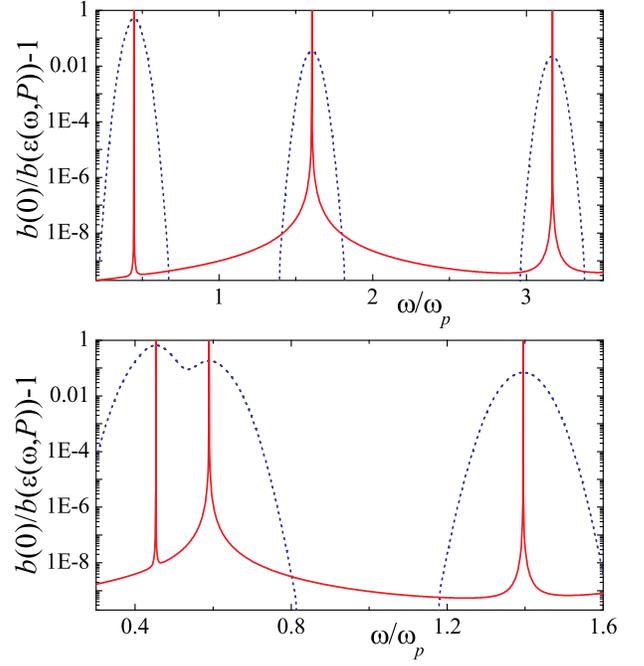}
\end{center}
\caption{\label{FigBomega} (Color online) The frequency dependence
of $b(0)/b(\bar{\varepsilon}(\omega,P))-1$, calculated for $d=2$
(upper graph) and $d=5$ (lower graph). Other parameters are
$I/I_c(d)=0.98$, $L=2\lambda_J$, and $v=1/30$ for both cases.
Solid curves correspond to monochromatic incoming radiation, while
the dashed curves describe the response on wide-band THz radiation
with $\bar{\gamma}=5\cdot10^{-2}$ (in units of $\omega_p$). The
radiation power $P$ is the same for all curves and is chosen such
that $\max(\bar{\varepsilon}(\omega,P))=0.3\varepsilon_0=2/45$ for
wide-band radiation.}
\end{figure}

Thus, we reduce the problem of quantum tunnelling of the field
$\varphi$ to the problem of tunnelling a quantum particle in two
dimensions, where the $\alpha_i$'s play the role of the particle
generalized ``coordinates''. The potential $V(\alpha_0,\alpha_1)$
is shown in Fig.~\ref{FigV}.

When $d>d_c(I,\bar{\varepsilon})\approx4$, there are three
solutions of the system of Eqs.~\eqref{eqAlpha} with the
conditions~\eqref{tauE}, $\alpha_i^{(0)}(\eta)$ and
$\alpha_i^{(\pm)}(\eta)$, which are characterized by the following
relations
$$
\alpha_1^{(0)}(\eta)=0\,,\,\,\,\,\alpha_1^{(-)}(\eta)=-\alpha_1^{(+)}(\eta)\,.
$$
The trajectories $\alpha_i^{(0)}(\eta)$ and
$\alpha_i^{(\pm)}(\eta)$ are shown in Fig.~\ref{FigV}. The
solution $\alpha_i^{(+)}(\eta)$ ($\alpha_i^{(-)}(\eta)$)
corresponds to the formation of vortex (antivortex) nucleus at
left (right) junction's edge, while the solution
$\alpha_i^{(0)}(\eta)$ describes the tunnelling of $\varphi$ as a
whole~\cite{weEPL}. The minimum of $B(\bar{E})$ corresponds to the
solutions $\alpha_i^{(\pm)}(\eta)$.  The tunnelling exponent then
reads
\begin{equation}\label{BEalpha}
B(\bar{E})=\frac{24{\Lambda}\mu_0^{5/2}}{5u_0^2}\,b(\bar{\varepsilon})\,,
\end{equation}
where
\begin{equation}\label{baplha}
b(\bar{\varepsilon})=\frac{15}{16}\int_{0}^{\eta_0}\!\!\!\!d\eta\left[\alpha_0^3+3\alpha_0\alpha_1^2-2\,\bar{\varepsilon}\right]\,,%
\,\,\,\,\eta_0=\sqrt{\mu_0}\tau_0\,.
\end{equation}
Note, that we should multiply the probability $\Gamma(\bar{E})$,
Eq.~\eqref{GammaE}, by a factor of $2$, since we have two channels
for tunnelling.

When $d<d_c(I,\bar{\varepsilon})$, all three solutions coincide,
$\alpha_1^{(0)}(\eta)=\alpha_1^{(-)}(\eta)=\alpha_1^{(+)}(\eta)=0$,
and the second equation of the system~\eqref{eqAlpha} becomes
trivial, while the first one can be easily integrated. As a
result, we obtain
\begin{equation}\label{baplha0}
b(\bar{\varepsilon})=\frac{15}{16}\int\limits_{\alpha_1(\bar{\varepsilon})}^{\alpha_2(\bar{\varepsilon})}\!\!\!\!d\alpha%
\frac{\alpha^3-2\bar{\varepsilon}}{\sqrt{\alpha^2(1-\alpha)-\bar{\varepsilon}}}\,,\,\,\,\,\,b(0)=1\,,
\end{equation}
where $\alpha_{1,2}(\bar{\varepsilon})$ are the smaller and larger
positive roots of the cubic equation
$$
\alpha^2(1-\alpha)-\bar{\varepsilon}=0\,.
$$

The analysis of the tunnelling exponent $B$ on the junction's
width $d$ and current $I$ was carried out in one~\cite{weEPL} of
our previous papers. Now we are interested in the effect of
electromagnetic radiation on $B(\bar{E})$. Using
formulas~\eqref{ELorenz}, \eqref{epsilon}, \eqref{baplha},
and~\eqref{baplha0}, we calculate the dependence of
$b(0)/b(\bar{\varepsilon}(\bar{\omega},P))-1$ as a function of
radiation's central frequency $\bar{\omega}$, for short $d<d_c$,
and long $d>d_c$ junctions. The results of the calculations, both
for wide-band ($\bar{\gamma}\gg\gamma_n$) and monochromatic
($\bar{\gamma}\ll\gamma_n$) radiation, are shown in
Fig.~\ref{FigBomega}. It is clear that we have several resonances
at frequencies $\bar{\omega}=\omega_n=\sqrt{\mu_n}$ (in units of
$\omega_{p}$). When $\bar{\gamma}=0$, the resonance peaks are very
narrow, and one can switch the JJ to the resistive state ($B=0$)
at small radiation power, if $\bar{\omega}=\omega_n$. Note, that
the condition $\bar{\omega}=\omega_n$ can be achieved by changing
the applied dc current $I$, since $\omega_n$ depend on $I$. In
other words, if the frequency of the incoming radiation lies near
one of the $\omega_n$, one can observe a resonance behavior of the
tunnelling exponent $B$ as a function of the dc current.

For relatively short junctions, $d\lesssim1$, we have several well
separated peaks even for wide-band THz radiation, as it can be
seen from Fig.~\ref{FigBomega}; while for $d\gtrsim d_c$, we have
$\omega_0\approx\omega_1\ll\omega_n$, and the first two peaks can
merge into a single peak. Note, that the inequality
$\omega_0\approx\omega_1\ll\omega_n$ is valid for not too high
junction's width $d\lesssim20$. In the opposite case we should
consider the large number of equations in the system~\eqref{eqC}
to calculate the tunnelling exponent.

\section{Conclusion}

In conclusion, we proposed a quantum field theory for Josephson
plasma waves interacting with external electromagnetic waves. We
also calculate the macroscopic quantum tunnelling of a fluxon,
stimulated by THz light, in a long Josephson junction driven by a
dc current. The probability of absorption and emission of light
depends on the current and the length of the Josephson junction.
The MQT escape rate shows several resonance maxima as a function
of the frequency, corresponding to eigenfrequencies of Josephson
plasma wave quanta. This could be potentially useful for a variety
of superconducting quantum THz devices. Classical THz devices are
discussed in Ref.~\onlinecite{LR}.

\section*{Acknowledgements}

We acknowledge partial support from JSPS-RFBR 06-02-91200.

FN and SS acknowledge partial support from Core-to-Core (CTC)
program supported by the Japan Society for Promotion of Science
(JSPS).

FN gratefully acknowledges partial support from the National
Security Agency (NSA), Laboratory Physical Science (LPS), Army
Research Office (ARO), National Science Foundation (NSF) grant No.
EIA-0130383.

SS acknowledges support from the Ministry of Science, Culture and
Sports of Japan via the Grant-in Aid for Young Scientists No
18740224, the UK EPSRC via No. EP/D072581/1, EP/F005482/1, and ESF
network-programme ``Arrays of Quantum Dots and Josephson
Junctions''.

AO acknowledges partial support from the Russian Foundation for
Basic Research (RFBR) (grant No. 06-02-16691), and Russian Science
Support Foundation. AO also acknowledges prof. A.\,L. Rakhmanov
for fruitful discussions.


\begin{thebibliography}{99}
\bibitem{animation}
Animations illustrating macroscopic quantum tunnelling in long
junctions can be found online in dml.riken.jp/MQT
\end{thebibliography}
\end{document}